# A Bidirectional Current-Mirror Based Potentiostat Using a Slice-Based Class-AB Amplifier

Markus Haberler, *Student Member, IEEE*, Inge Siegl, Christoph Steffan, and Mario Auer, *Member, IEEE*

*Abstract*—A pico-ampere sensitive electrochemical sensor signal acquisition chain consisting of a multi-channel bidirectional current-mirror based potentiostat and a multiplexed current-input delta-sigma ($\Delta\Sigma$) ADC is presented. Slice-based class-AB amplifiers with output stage current replication using reconfigurable current-mirrors enable bidirectional current sensing over a range of five decades. The replicated current is digitized by the ADC. The proposed circuit achieves a minimum detectable current of $91.7\,\text{pA}$, resulting in a dynamic range of $147\,\text{dB}$ for sensor currents up to $\pm 1\,\text{mA}$. The potentiostat is operated at a $3\,\text{V}$ supply, the ADC at $1.5\,\text{V}$. The static power consumption of a single interface channel depends on the number of active amplifier slices and is between $14.1\,\mu\text{W}$ and $39.3\,\mu\text{W}$. The highest power efficiency among similar systems and the low voltage headroom consumption of less than $500\,\text{mV}$ enables a versatile use of the interface for multiple sensing applications.

## I. Introduction

In order to improve the medical situation in rural areas, fast and inexpensive diagnostics are of great importance. Point-of-care (PoC) diagnostics promise to replace the need for specialized equipment by offering rapid and accurate test results while keeping the medical costs at a minimum [1]. The integration of biochemical functionalities with a performance comparable to laboratory-based equipment, combined with wireless operation, are key aspects to enable widespread adoption of PoC devices.

To analyze various biochemical parameters, different electrochemical sensing methods including constant potential amperometry (CA), cyclic voltammetry (CV) and electrochemical impedance spectroscopy (EIS) have to be provided. These analyses require a controlled-potential sensor interface that is able to process bidirectional (reduction and oxidation) sensor currents. The limited available power in wireless operation requires a low power and low voltage headroom consuming implementation of the corresponding circuits.

A PoC diagnostic platform that uses a two-chip solution for this purpose was recently presented [2]. One chip is used to perform the desired electrochemical measurements, the other is used for data acquisition and data transfer. Additionally, an external energy storage device is used to power the diagnostic

Manuscript received May 14, 2020. This research is part of a project that has received funding from the European Union's Horizon 2020 research and innovation program under grant agreement No 761167 (IMPETUS).
M. Haberler, I. Siegl, and C. Steffan are with Infineon Technologies Austria AG, Graz, Austria (e-mail: markus.haberler@infineon.com).
M. Auer is with Graz University of Technology, Graz, Austria (e-mail: mario.auer@tugraz.at).

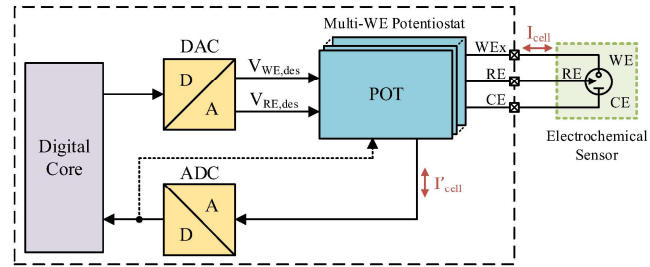

Fig. 1. Block diagram of the controlled-potential biochemical sensor interface. The DAC provides the excitation signal to the sensor front-end that consists of a multi-WE potentiostat. The potentiostat drives the electrodes and generates a sensor current replica that is digitized by the ADC and stored on-chip. The digitized information is utilized to reconfigure the potentiostat.

platform. A battery-less single-chip solution was presented in [3]. The platform uses the electromagnetic field for power supply and data transfer. A low voltage topology of the sensor front-end is used; however, the front-end is only capable of processing unidirectional sensor currents. This limits the usability of the system to analyses that obtain the information from either reduction or oxidation currents. To address this problem, an adaption of the interface circuit was performed in [4]. Bidirectional currents are possible, provided by a manual selection of the expected current direction. However, this has only limited suitability for accurate dynamic measurements as required in CV. Several other works also addressed the problem of bidirectional current sensing using various techniques. The solution presented in [5] requires manual adaption. In [6] offset currents are added, which influence the accuracy and the power consumption. A topology that consumes considerable voltage headroom is proposed in [7].

In this letter, we present a sensor interface circuitry which is able to process bidirectional sensor currents without additional measures, while keeping the voltage headroom consumption at a minimum. This is achieved by a signal chain consisting of a class-AB amplifier based sensor front-end followed by a delta-sigma ($\Delta\Sigma$) ADC. The low power and area consumption allows the circuit to be integrated into a single-chip PoC diagnostic platform that uses the electromagnetic field for power supply and data transfer.

## II. Controlled-Potential Sensor Interface

Electrochemical analyses like CA, CV or EIS are usually carried out by a controlled-potential sensor interface. A block diagram of the interface used in this work is shown in Fig. 1. The controlled-potential interface is based on a three-electrode potentiostat that sets the voltage across the electrochemical sensor by two operational amplifiers that are operated in a feedback-configuration. Depending on the desired electrochemical analysis, the potentiostat maintains a potential difference between the working electrode (WE) and the reference electrode (RE) by adjusting the potential of the counter electrode (CE). The current that arises due to the electrochemical reaction is measured [8].

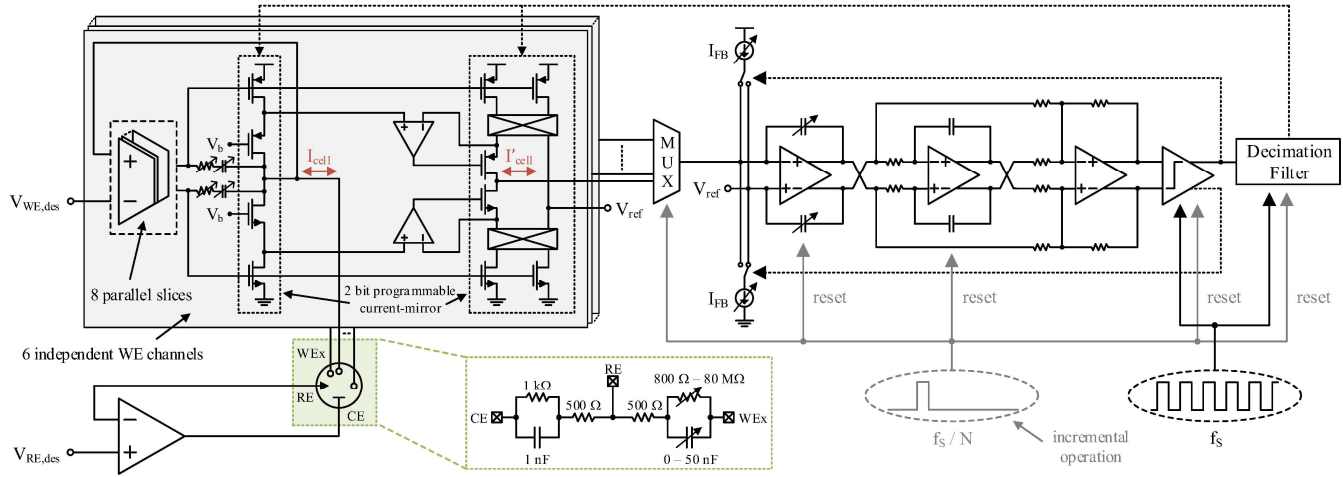

Fig. 2. Sensor signal acquisition chain of the multi-WE controlled-potential sensor interface. The amplifier used to control the WE consists of a folded-cascode amplifier with a class-AB output stage. The slice-based implementation of the first gain stage ensures stability for strong varying load conditions. Programmable current-mirrors are used to generate a scaled version of the sensor current. Regulated-cascodes and DEM techniques are used to improve the accuracy. The current-input $\Delta\Sigma$ ADC is multiplexed between the WE channels and processes the scaled sensor current.

To simultaneously determine several biochemical key figures, multiple sensor interface channels are required that control differently composed electrodes. To induce the redox reaction for the target biomolecules, potential differences of up to $2.5\,\mathrm{V}$ are necessary at the sensor electrodes. Together with the wireless operation of the platform, stringent requirements on the individual circuit blocks of the sensor interface result.

The implemented multi-WE sensor front-end consists of bidirectional current-mirror based potentiostats that avoid a voltage headroom consuming current-to-voltage transformation, thus allowing proper operation with a supply voltage as low as $3\,\mathrm{V}$. Reconfigurable current-mirrors combined with the application of matching improvement techniques provide a wide dynamic range in the potentiostats. Sensor current and sensor properties dependent frequency compensation techniques enable a versatile use of the interface for multiple sensing applications. A detailed description of the implemented potentiostat is given in Section III. A current-input $\Delta\Sigma$ ADC is used to process the sensor current replica provided by the potentiostat. As will be discussed in Section IV, incremental operation of the ADC allows multiplexing between the multiple WE channels, thus saving power and area.

## III. BIDIRECTIONAL CURRENT-MIRROR BASED POTENTIOSTAT

Each channel of the potentiostat is required to process bidirectional sensor currents over a dynamic range of five decades, while consuming a voltage headroom of only $500\,\mathrm{mV}$. To achieve the desired accuracy, the current error has to be less than $2\,\%$. In order to enable wireless operation, the current consumption of the circuit has to be kept at a minimum. The topology implemented for this purpose is shown in Fig. 2.

The proposed circuit draws on the well-known topology of current-mirror based potentiostats [3], extended for bidirectional sensor currents. This is achieved by using a class-AB output stage in the WE controlling amplifier that replicates the currents through both output stage transistors. These transistors are driven by a folded-cascode amplifier that uses floating class-AB control transistors [9]. Since the bidirectional current-mirror relies on the subtraction of two currents, a low quiescent current in the output stage is required to achieve the desired current accuracy [10]. This is achieved by suitable design of the floating class-AB control transistors and the output stage transistors.

The use of programmable current-mirrors that enable a ratio dependent optimization based on current scaling ensure the desired accuracy over the wide dynamic range with minimum area and voltage headroom consumption [11]. The potentiostat is able to process bidirectional currents in a range of $10\,\mathrm{nA}$ to $1\,\mathrm{mA}$. The 2-bit programmable current-mirrors with scaling factors of 1, 10, 100, and 1 000, allow a downscaling of the sensor current to a range of $1\,\mu\mathrm{A}$ to $1\,\mu\mathrm{A}$.

Regulated-cascodes in the current-mirrors are used to minimize errors related to the finite output impedance of the transistors. Proper operation of both regulated-cascodes is ensured by adding transistors with a constant bias voltage in the class-AB output stage. Depending on the applied WE potential, these transistors operate either in saturation or triode region, thereby generating the required voltage shifts for the regulated-cascodes. These voltage shifts additionally allow the use of low-voltage, thin gate oxide transistors in the current-mirrors, which further improve the current processing accuracy due to better matching properties. The utilization of dynamic element matching (DEM) techniques results in an additional accuracy improvement in the current-mirrors. Depending on the sensing application, either manual switching or chopping at a frequency of $300\,\text{kHz}$ can be performed for the mirror side transistors.

The wide dynamic range of sensor currents and the thereby associated strong varying load conditions require an adaptive approach for the frequency compensation to ensure stability in the feedback system. A dynamically scalable folded-cascode amplifier enables a sensor current dependent shifting of the first stage pole without changing the gain of the stage. This is achieved by a slice-based implementation of the first stage of the amplifier which is controlled by 3 bits. Together with an adjustable Miller compensation, stability for sensors with an ohmic and ohmic-capacitive behavior is achieved.

## IV. Delta-Sigma ADC Based Current Readout

The potentiostat provides a downscaled version of the sensor current in the range of $1\,\text{µA}$ to $1\,\text{µA}$. For an accurate biochemical analysis, this current needs to be digitized with a resolution of $500\,\text{pA}$. Hence, the required resolution of the ADC is roughly 12 bits. Furthermore, a signal bandwidth of up to $10\,\text{kHz}$ is necessary to perform EIS analyses. A current-to-frequency (I-to-F) converter was used in [3], whereas a dual-slope topology was used in [4]. Without manual adaption, both of these ADC types are again just able to process unidirectional sensor currents.

In this work a current-input $\Delta\Sigma$ modulator is used. A second-order continuous-time loop filter in a feedforward topology is implemented, as shown in Fig. 2. An oversampling ratio of 120 is used to achieve the desired resolution, resulting in a sampling frequency of $2.4\,\text{MHz}$. A current-steering feedback DAC is used to apply the feedback signal. Due to the closed loop operation of $\Delta\Sigma$ modulators, direct processing of bidirectional sensor currents is possible.

A selection of the input full-scale range can be performed using a 4-bit adjustable current DAC and feedback capacitor in the first integrator. This allows to increase the current resolution for applications with low sensor currents.

Since the expected biochemical sensor signals are in the low frequency range, flicker noise is the main limiting factor in the design. As a countermeasure, source degeneration in the current-steering feedback DAC is applied, which reduces the flicker noise of the implementation by more than $6\,\text{dB}$.

Due to the limited available power in wireless operation, the ADC is designed to operate down to $1.5\,\text{V}$. This is achieved by independently selectable input and output common-mode voltage levels of the input integrator. Feedforward-compensated amplifiers are used in the loop filter due to their high power efficiency [12]. Furthermore, only one multiplexed ADC is used for the multiple sensor interface channels. This is realized by operating the $\Delta\Sigma$ ADC in incremental mode in case a multi-WE analysis is performed.

The digitized sensor current is stored in the on-chip RAM and prepared for wireless data transmission. Additionally, the information of the sensor current is used to reconfigure the potentiostat to achieve optimum performance.

## V. Measurement Results and Evaluation

The transfer characteristic of the bidirectional potentiostat and the error of the current processing are shown in Fig. 3.

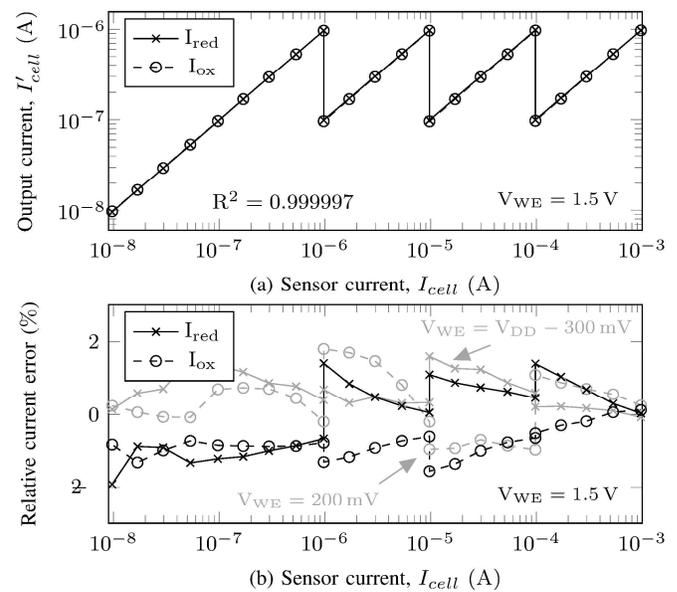

Fig. 3. Measured performance of the bidirectional current-mirror based potentiostat. (a) Transfer characteristic of the potentiostat. (b) Relative current error of the potentiostat.

In these plots, the solid traces show the current processing performance for currents that flow from WE to CE (reduction currents), whereas the dashed traces show the performance for currents that flow from CE to WE (oxidation currents). Similar performance in both directions is achieved with a maximum error of less than $2\,\%$, verified for a set of 10 randomly chosen dies. The circuit exhibits high linearity with an $R^2$ value of $0.999997$. The voltage headroom consumption of the potentiostat is less than $500\,\text{mV}$ over the complete dynamic range. The current consumption of a single channel of the potentiostat depends on the number of active slices of the folded-cascode amplifier. Each slice consumes $1.2\,\text{µA}$, which leads to a maximum static power consumption of less than $29\,\text{µW}$ at a nominal supply of $3\,\text{V}$. The area consumption of a single channel of potentiostat is approximately $0.11\,\text{mm}^2$.

The power spectral density (PSD) of the $\Delta\Sigma$ modulator for a sinusoidal input current with an amplitude of $500\,\text{nA}$ is shown in Fig. 4. The achieved SNDR in a signal band of $10\,\text{kHz}$ is higher than $73.3\,\text{dB}$, which results in an ENOB of almost 12 bits. Hence, with nominal settings the scaled sensor current can be digitized with a resolution of $530\,\text{pA}$. By adjusting the full-scale range of the ADC, the minimum detectable current can be reduced to $91.7\,\text{pA}$. The circuit consumes $30\,\text{µA}$ from a $3\,\text{V}$ supply. Reducing the supply voltage to $1.5\,\text{V}$ results in a reduction of the power consumption by a factor of 2, without affecting the performance of the modulator. This results in a Schreier FOM of $156.8\,\text{dB}$. The area consumption of the modulator is only $0.05\,\text{mm}^2$.

The circuit was realized in a $130\,\text{nm}$ standard CMOS process as a main building block of a wirelessly-operated electrochemical measurement platform IC. Fig. 5 shows a micrograph of the presented circuits. The IC contains the multi-WE potentiostat consisting of six equal implementations of the presented bidirectional current-mirror based topology.

TABLE I
COMPARISON OF SELECTED IMPLEMENTATIONS

|  | This work | [2] | [3] | [4][a] | [5] |
|---|---|---|---|---|---|
| Technology (μm) | 0.13 | 0.18 | 0.18 | 0.35 | 0.18 |
| Sensor Current Range (A) | ±10 n to ±1 m | ±200 μ | +1 n to +1 μ | ±p to ±μ | ±18 μ |
| Resolution | 91.7 pA @ ±100 nA | 94 pA @ ±2 μA | <1 nA | - | 0.9 pA |
| Dynamic Range (dB) | 147 | 133 | >60 | - | 152 |
| Supply Voltage (V) | 3 / 1.5 | 2.5 | 1.8 | 1.2 | 1.5 |
| Voltage Headroom Consumption (V) | <0.5 | <1.5 | - | - | - |
| Static Power Consumption / Ch. (μW)[b] | 14.1–39.3 | 3250 | 50.4 | 12.1 | 2.9 |
| Power Efficiency[c] | 0.66 | 0.12 | 0.02 | - | 0.59 |
| Bidirectional Currents | yes | yes | no | yes[d] | yes[d] |
| Sensing Methods (CA/CV/EIS) | yes/yes/yes | yes/yes/- | yes/-/- | yes/yes[d]/- | yes/yes[d]/- |
| ADC type | delta-sigma | - | I-to-F | dual-slope | I-to-F |

[a] Based on simulation    [b] Sensor current excluded    [c] Power Eff. = max. current range / dyn. power consumption    [d] Based on manual adaption

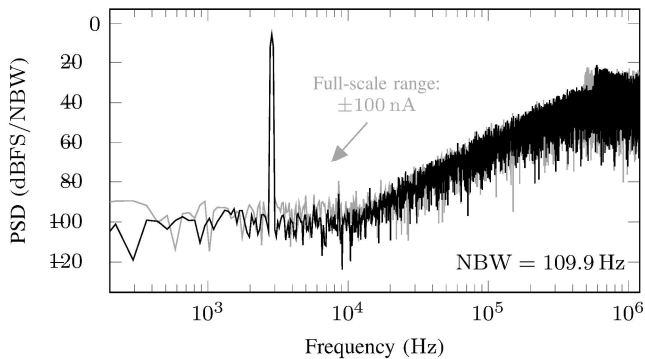

Fig. 4. Output PSDs of the current-input $\Delta\Sigma$ modulator. In nominal configuration the SNDR in the frequency band of 10 kHz is higher than 73 dB.

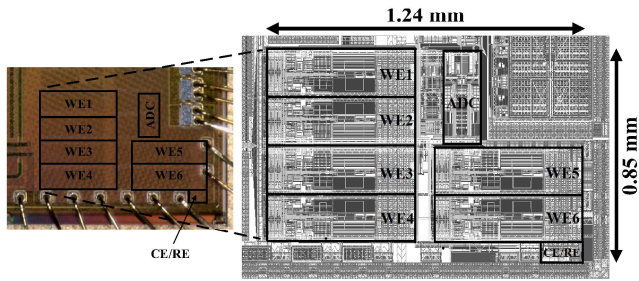

Fig. 5. Die micrograph of the multi-WE sensor interface.

The ADC is multiplexed between the individual channels. The average static power consumption per interface channel is between $14.1\,\mu W$ and $39.3\,\mu W$. Due to the class-AB amplifier based front-end topology, the power consumption during current sensing is mainly determined by the sensor current itself. This leads to a maximum power consumption of only 3 mW while sensing a current of 1 mA, resulting in a power efficiency [5] of 0.66. The active area consumption of the interface is approximately $0.7\,mm^2$.

A comparison of the presented design to a selection of existing biochemical signal acquisition chains is provided in Table I. The proposed solution achieves the highest power efficiency and a wide dynamic range. The low voltage headroom consumption together with the ability to process bidirectional sensor currents enables the circuit to induce the redox reaction for multiple sensing applications.

## VI. CONCLUSION

A power-efficient bidirectional sensor signal acquisition chain was presented. The proposed circuit, which consists of a potentiostat with output stage current replication and $\Delta\Sigma$ ADC based current processing, enables wide dynamic range current sensing with minimal voltage headroom consumption. The class-AB topology of the front-end amplifier keeps the current consumption at a minimum, which results in the highest power efficiency of the implementation when compared to state-of-the-art solutions. These properties make the biochemical measurement platform broadly applicable as a low-cost solution for wirelessly-powered PoC diagnostics.